\documentclass{article}
\usepackage[dvips]{graphicx}
\begin{document}
\title{Chaotic diffusion of particles with finite mass in oscillating convection flows}
\author{Hidetsugu Sakaguchi\\
Department of Applied Science for Electronics and Materials,\\ Interdisciplinary Graduate School of Engineering Sciences,\\
 Kyushu University, Kasuga, Fukuoka 816-8580, Japan}
\maketitle
\begin{abstract}
Deterministic diffusion in temporally oscillating convection is studied for particles with finite mass.  The particles are assumed to obey a simple dissipative  dynamical system and the particle diffusion is induced by the strange attractor.  The diffusion constants are numerically calculated for convection models with free and rigid boundary conditions. 
 \\
\\
PACS numbers:05.40.-a, 05.45.-a, 47.52.+j
\end{abstract}
\newpage
The motion of a fluid particle in the velocity field  ${\bf v}({\bf x})$ 
is determined by the differential equation.
\begin{equation}
\frac{d{\bf x}}{dt}={\bf v}({\bf x},t)
\end{equation}
with the initial condition ${\bf x}(0)$.
Even for regular velocity fields, the particle motion can become 
chaotic and it is called the Lagrangian chaos\cite{rf:1,rf:2}. For an incompressible fluid, the dynamical system (1) is a conservative system.
In two dimensions, the evolution equations become
\begin{equation}
\frac{dx}{dt}=\frac{\partial \psi}{\partial z},\;\;\frac{dz}{dt}=-\frac{\partial \psi}{\partial x},
\end {equation} 
where $\psi$ is the stream function and ${\bf x}=(x,z)$. Equation (2) is a Hamiltonian system.  Solomon and Gollub studied experimentally and numerically the particle diffusion in Rayleigh-B\'enard convection which oscillates temporally \cite{rf:3}.  
In the oscillating convection, the velocity field is derived from a stream function $\psi=A/k\sin\{kx+B\sin(\omega t)\}W(z)$, where $k$ is the wavenumber, $A$ is the maximum vertical velocity, and $B$ and $\omega$ are the amplitude and the frequency of the oscillation and $W(z)$ is a function which satisfies the boundary conditions at the top and bottom surfaces ($z=1$ and 0).

In this report, we assume that a particle embedded in the fluid has a finite mass. The equation of motion is in general rather complicated \cite{rf:4}.
We assume a simple model as
\begin{equation}
\frac{dv_{x}}{dt}=\gamma\left(\frac{\partial \psi}{\partial z}-v_x\right),\;\;\frac{dv_z}{dt}=\gamma\left(-\frac{\partial \psi}{\partial x}-v_z\right),
\end{equation} 
where $v_x=dx/dt, v_z=dz/dt$, and $\gamma^{-1}$ denotes a response time and is proportial to the mass of the particle. For the massless case, $\gamma$ becomes infinity and Eq.~(2) is recovered. The volume contraction rate of the four dimensional phase space $(x,v_x,z,v_z)$ is $-2\gamma$, and therefore the dynamical system (3) is a dissipative system. We assume the same form of stream function $\psi=A/\pi\sin\{\pi x+B\sin(2\pi t)\}W(z)$.  For simplicity, we assume $W(z)=\sin(\pi z)$ for the convection with free boundary conditions, and $W(z)=z^2(1-z)^2$ for the convection with rigid boundary conditions. More complicated function $W(z)$ is used in the exact linear stability analysis of the B\'enard convection with  rigid boundary conditions \cite{rf:5}. 

The fixed points of the dynamical system (3) at $B=0$  are $(x_0,v_{x0},z_0,v_{z0})=(1/2,0,1/2,0)$, and $(x_1,v_{x1},z_1,v_{z1})=(0,0,0,0)$ and $(x_2,v_{x2},z_2,v_{z2})=(0,0,1,0)$. The fixed point $(x_0,z_0)$ is an elliptic point for the Hamiltonian system (2) and the orbit near the elliptic point is stable for small perturbation in the Hamiltonian system.  It is called the KAM theorem and the chaotic behaviors cannot be expected near the elliptic point. The fixed points $(x_1,z_1)$ and $(x_2,z_2)$ are saddle points and the chaotic orbits appear near the saddle points for nonzero $B$ in the Hamiltonian system. 

On the other hand, 
the fixed point $(x_0,v_{x0},z_0,v_{z0})$ for Eq.~(3) at $B=0$  is an unstable focus for finite $\gamma$ and the orbit approaches the heteroclinic orbits which connects the different saddle points. Figure 1(a) displays a trajectory starting from $(x,v_{x},z,v_z)=(0.5001,0,0.5,0)$ for $A=1, B=0, \gamma=20$ and $W(z)=\sin\pi z$.  As the time evolves, all trajectories approach 
the saddle connections which connect four saddle points $(0,0),(1,0),(1,1)$ and $(1,0)$.  The numerical simulation was performed with the Runge-Kutta method of timestep $\Delta t=0.0005$. The orbits seem to be attracted rapidly to the saddle points.
\begin{figure}[htb]
\begin{center}
\includegraphics[width=8cm]{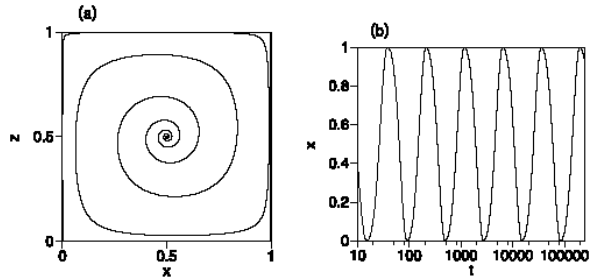}
\caption{(a) Trajectory in the $(x,z)$ plane for $A=1,B=0,\gamma=20$ and $W(z)=\sin\pi z$. The initial condition is $(x,v_x,z,v_z)=(0.5001,0,0.5,0)$.
(b) Time evolution of $x(t)$ for $A=4\pi,B=0,\gamma=20$ and $W(z)=z^2(1-z)^2$. The initial condition is $(x,v_x,z,v_z)=(0.1,0,0.1,0)$. Note that the time scale is logarithmic. 
} 
\label{fig:1} 
\end{center}
\end{figure}
For the rigid boundary model with $W(z)=z^2(1-z)^2$, there are fixed lines at $z=0$ and 1.   
The eigenvalues around the points $(x,z)=(x,0)$ and $(x,1)$ on the fixed lines are calculated from 
\[
\frac{d\delta x}{dt}=\delta v_x,\;\;\frac{d\delta v_x}{dt}=\gamma(2A/\pi\sin(\pi x)\delta z- \delta v_x) ,\;\;\frac{d\delta z}{dt}=\delta v_z,
\;\;\frac{d\delta v_z}{dt}=-\gamma \delta v_z,\]
where $\delta x,\delta z,\delta v_x,\delta v_z$ are small deviations from the points on the fixed lines.  The eigenvalues are $0,0,-\gamma$ and $-\gamma$. It is a singular situation 
in a  viewpoint of dynamical systems.
The trajectories are attracted to the saddle connection, however, the flow near the fixed lines  is very slow. Figure 1(b) displays the time evolution of $x(t)$  starting from $(x,v_{x},z,v_z)=(0.1,0,0.1,0)$ for $A=4\pi, B=0, \gamma=20$ and $W(z)=z^2(1-z)^2$. The time evolution of $x(t)$ may seem to be like a limit cycle, however, the time axis is plotted with a logarithmic scale. It implies 
that it takes exponentially longer time for the particle to circulate one convection cell as the particle is close to the saddle connection.

The saddle connection is unstable for general perturbations. 
The unstable manifolds and stable manifolds starting from  saddle points 
intersect and 
a strange attractor appears near the saddle points for nonzero $B$ \cite{rf:6}. All trajectories are attracted  to the strange attractor near the saddle points and the chaotic diffusion is induced.  
\begin{figure}[htb]
\begin{center}
\includegraphics[width=12cm]{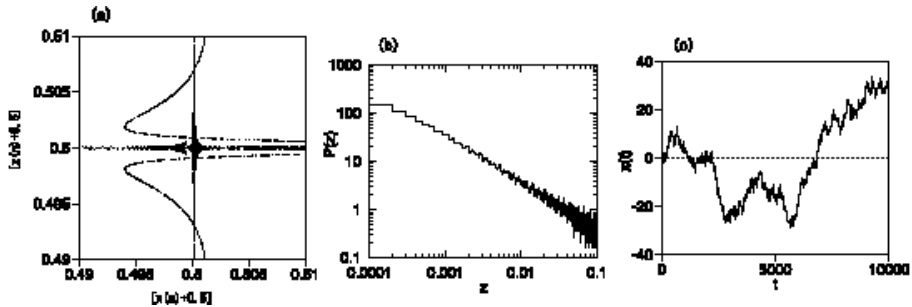}
\caption{(a) Strange attractor in the $([x+1/2],[z+1/2])$ plane for $A=1,B=0.001,\gamma=20$ and $W(z)=\sin\pi z$. (b) Stationary distribution $P(z)$ of $z$ for the same parameter values as (a).(c) Time evolution of $x(t)$ for the same parameters.
} 
\label{fig:2} 
\end{center}
\end{figure}
Figure 2(a) is a projection of the strange attractor
 into $([x+1/2],[z+1/2])$ plane at $t=n=$integer for $A=1,B=0.001,\gamma=20$ and $W(z)=\sin\pi z$, where $[y]$ implies the fractional part of $y$.
The strange attractor around the saddle points $(x,z)=(0,0),(1,0),(0,1)$ and $(1,1)$ is shifted around $(1/2,1/2)$ to clarify the attractor.  
We have performed several numerical simulations using different initial conditions, however, we have obtained the same strange attractor.   
The stationary distribution of the particle position is uniform for the Hamiltonian system (2) which corresponds to the massless case, however, the stationary distribution of the particle position is not uniform in the dissipative system (3).  Figure 2(b) displays a double logarithmic plot of the distribution $P(z)$ of $z$ near $z=0$ for $A=1,B=0.001,\gamma=20$ and $W(z)=\sin\pi z$.  The particle is attracted to the strange attractor near the saddle points and the stationary distribution  seems to obey a power law  with exponent about 0.93. 
Figure 2(c) displays the time evolution of $x(t)$ for the same parameters from the initial condition $(x,v_x,z,v_z)=(0.5001,0,0.5,0)$.
Chaotic diffusion is induced even for the very small perturbation.

There are many "windows" of stable limit cycle attractors when the parameter 
$B$ is changed. If a limit cycle which runs across convection cells 
exists and it is stable, $x(t)$ increases or decreases with a constant average velocity. The limit cycle corresponds to the accelerator mode in the Hamiltonian  system (2) \cite{rf:7}.  As $B$ is changed, the stable limit cycle disappears at a critical point. However, the velocity correlation is very long near the bifurcation point of the limit cycle.  The velocity $v_x$ keeps positive or negative values for a long time and the direction of motion changes intermittently.
Figure 3 displays such a constantly increasing time evolution of $x(t)$ at $B=0.002$ and an intermittent time evolution of $x(t)$ at $B=0.002165$ for $A=1,\gamma=20$ and $W(z)=\sin\pi z$.   
\begin{figure}[htb]
\begin{center}
\includegraphics[width=8cm]{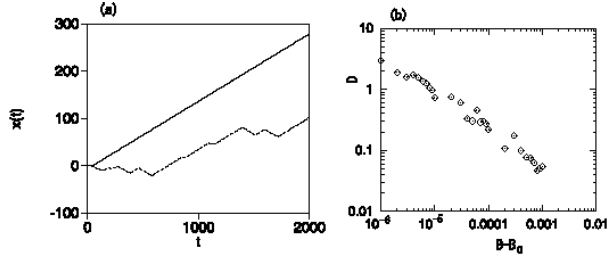}
\caption{(a) Time evolution $x(t)$ at $B=0.002$ (solid line) and 0.002165 (dashed line) for $A=1, \gamma=20$ and $W(z)=\sin \pi z$.
(b) Diffusion constants for several $B$'s near the critical point of the limit cycle.} 
\label{fig:3} 
\end{center}
\end{figure}
The diffusion constant $D=\lim_{\tau\rightarrow \infty}\langle(x(t+\tau)-x(t))^2\rangle/2\tau$ is enhanced 
near the bifurcation point.  Figure 3(b) displays the diffusion constant as a function of $B$. The average $\langle \cdots \rangle$ to evaluate the diffusion constant is numerically calculated as double averages of a long time average and the average with respect to 100 different initial conditions. The diffusion constant increases as $1/(B-B_c)^{\alpha}$, where $B_c\sim0.00216$ and $\alpha=0.5\sim 0.6$.

The chaotic diffusion occurs both for free boundary models and rigid boundary models. 
The difference between the two boundary conditions is clearly seen, when the perturbation amplitude $B$ is very small.  
Figure 4 displays the averge period for a particle to circulate one convection cell as a function of $B$. The period is numerically obtained as the average interval between neighboring two times where $z(t)=0.5$ and $dz/dt>0$. 
Figure 4(a) shows the results for $A=1,\gamma=20, W(z)=\sin\pi z$.  The average period $T$ is approximately $T\sim -4.1\log_{10}B$.
The logarithmic dependence is characteristic of the saddle connection. 
The period  increases very slowly as $B$ is decreased.
Figure 4(b) shows the results for $A=4\pi, \gamma=20$ and $W(z)=z^2(1-z)^2$.
\begin{figure}[htb]
\begin{center}
\includegraphics[width=8cm]{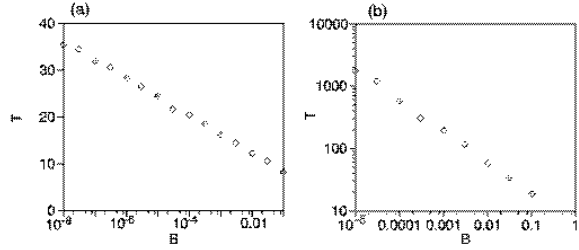}
\caption{(a) Semi-logarithmic plot of the average  circulation period for $W(z)=\sin\pi z$. (b) Double-logarithmic plot of the average circulation period for $W(z)=z^2(1-z)^2$.} 
\label{fig:4} 
\end{center}
\end{figure}
The average period increases approximately by a power law: $T\sim B^{-1/2}$ as 
$B$ is decreased for the rigid boundary model. It may be due to the existence of the fixed lines for the rigid boundary conditions. 
The average period $T$ plays a role of a unit time for the chaotic diffusion. 
For example, if the direction of motion is assumed to be randomly changed  after each circulation of convection cells, the total steps $N$ of random changes of direction during the time interval $\tau$ is estimated as $N\sim \tau/T$, therefore, the diffusion constant is 
estimated as $D\propto 1/T$.
Figure 5(a) displays numerically obtained diffusion constants at several $B$'s for $A=1,\gamma=20, W(z)=\sin\pi z$. The diffusion constants depend on $B$ in a complicated manner. It is partly due to the "windows" structures by the stable limit cycles. For example, the diffusion constant $D$ is zero at $B=3\times 10^{-5}$, where there exists a stable limit cycle which circulates in a single convection cell. However, 
the diffusion constant has a tendency to decrease very slowly as $B$ is decreased. Figure 5(b) displays numerically obtained diffusion constants at several $B$'s for $A=4\pi, \gamma=20$ and $W(z)=z^2(1-z)^2$. The diffusion constant decreases with a power law: $D\sim B^{1/2}$ as $B$ is decreased. 
The numerical results are consistent with the above rough argument.
\begin{figure}[htb]
\begin{center}
\includegraphics[width=8cm]{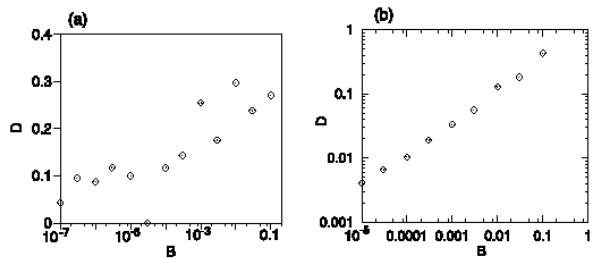}
\caption{(a) Semi-logarithmic plot of the diffusion constants for $W(z)=\sin\pi z$. (b) Double-logarithmic plot of the diffusion constants for $W(z)=z^2(1-z)^2$.} 
\label{fig:5} 
\end{center}
\end{figure}

To summarize, we have numerically studied the chaotic diffusion of particles with finite mass in oscillating convection.  
The  diffusion is induced by the strange attractor near the saddle points. 
In a sence, the chaotic diffusion is enhanced in such dissipative systems, since there exist no stable oribits near the center of the convection cell and the orbits are attracted toward the saddle points in contrast to the Hamiltonian system. 
The strange attractor exists even for very small perturbations, since the 
attractor of the nonperturbed system is a singular one, that is, the saddle-connection. 
Special enhancement of the diffusion constant occurs near the bifurcation points of  limit cycles corresponding to the accelerator modes in the Hamiltonian system. 
The behaviors of the chaotic diffusion at very weak $B$ depend on the boundary conditions. For the rigid boundary conditions, the motion of the particle becomes very slow near the top and bottom boundaries, since the flow velocity is zero  at the boundaries. Therefore, the average period circulating around one cell becomes very large for small $B$ and the diffusion constant decreases more rapidly than the case of the free boundary conditions.  
However, it is left to future study to explain the $B$ dependence theoretically. 

We would like to thank Professor S.Kai, Dr.Y.Hidaka and Mr.K.Tamura for valuable discussions.

\end{document}